# Analyzing signal attenuation in PFG anomalous diffusion via a non-Gaussian phase distribution approximation approach by fractional derivatives

Guoxing Lin[a)]

*Carlson School of Chemistry and Biochemistry, Clark University, Worcester, MA 01610*

**Abstract**

Anomalous diffusion exists widely in polymer and biological systems. Pulsed field gradient (PFG) techniques have been increasingly used to study anomalous diffusion in NMR and MRI. However, the interpretation of PFG anomalous diffusion is complicated. Moreover, there is not an exact signal attenuation expression based on fractional derivatives for PFG anomalous diffusion, which includes the finite gradient pulse width effect. In this paper, a new method, a Mainardi-Luchko-Pagnini (MLP) phase distribution approximation, is proposed to describe PFG fractional diffusion. MLP phase distribution is a non-Gaussian phase distribution. From the fractional diffusion equation based on fractional derivatives in both real space and phase space, the obtained probability distribution function is a MLP distribution. The MLP distribution leads to a Mittag-Leffler function based PFG signal attenuation rather than the exponential or stretched exponential attenuation that is obtained from a Gaussian phase distribution (GPD) under a short gradient pulse approximation. The MLP phase distribution approximation is employed to get a complete signal attenuation expression $E_\alpha(-D_f b^*_{\alpha,\beta})$ that includes the finite gradient pulse width effect for all three types of PFG fractional diffusion. The results obtained in this study are in agreement with the results from the literature. These results provide a new, convenient approximation formalism to interpret complex PFG fractional diffusion experiments.

[a)]Electronic mail:glin@clarku.edu

## I. INTRODUCTION

The pulsed field gradient (PFG) diffusion technique[1,2,3] has been used to measure anomalous diffusion in NMR and MRI. Anomalous diffusion exists in many systems such as in polymer or biological systems, porous materials and single-file structure.[4,5,6,7,8,9,10,11] Based on continuous-time random walk (CTRW), anomalous diffusion arises from its diffusion waiting time distribution behaving asymptotically to a power law related to time derivative order $\alpha$, or from its jump length distribution following a power law[6,12] pertinent to space derivative order $\beta$. The interpretation of PFG anomalous diffusion is relatively complicated compared to normal diffusion.[13] Unlike normal diffusion, the anomalous diffusion usually has a mean square displacement not proportional to diffusion time, and may have a non-Gaussian distribution function.[14,15] Some types of anomalous diffusion can be modeled by the fractional diffusion (FD) equation based on the fractional derivatives (see Appendix A)[9,14,15]. Though there is much effort to develop theoretical treatments for anomalous diffusion[16,17,18,19,20,21,22,23,24,25,26,27,28,29], many challenges remain in interpreting PFG anomalous diffusion, most notably the finite gradient pulse width (the signal attenuation during the gradient pulse applying period) effect.

The Gaussian phase distribution (GPD) approximation has been employed by Kärger et al. to analyze PFG anomalous diffusion, which successfully obtained a stretched exponential signal attenuation.[16] However, the results[16] only include the time fractional diffusion, which is one of the three types of fractional diffusions. Based on the values of time derivative order $\alpha$ and space derivative order $\beta$, the fractional diffusions can be divided into general fractional diffusion $\{0 < \alpha, \beta \leq 2\}$, time fractional diffusion $\{0 < \alpha \leq 2, \beta = 2\}$, and space fractional diffusion $\{\alpha = 1, 0 < \beta \leq 2\}$.[14,28] Additionally, the accumulating phase shift (APS) distribution in the PFG experiment has been found not to be a GPD for fractional diffusion based on fractional derivatives.[28] Based on the results of the recently developed effective phase shift diffusion equation (EPSDE) method,[28] the function of APS distribution is indeed a probability distribution function (PDF) described as $(D_{\phi eff} t^\alpha)^{-1/\beta} \kappa_{\beta,\alpha}^\theta \left( \frac{\phi}{(D_{\phi eff} t^\alpha)^{1/\beta}} \right)$ [14,15,28,30] where $D_{\phi eff}$ is the effective phase fractional diffusion coefficient with units of rad$^\beta$/s$^\alpha$, $\theta$ is the skewness, $t$ is time, and the $\kappa_{\beta,\alpha}^\theta(x)$ function is defined in Appendix B. This phase distribution will be referred to as "Mainardi-Luchko-Pagnini phase distribution (MLPPD)".[14,15,28,30] Moreover, under the short gradient pulse (SGP) approximation, such an MLPPD leads to a Mittag-Leffler function (MLF) (see Appendix B) signal attenuation,[14,25,28] which differs from the stretched exponential function attenuation obtained in references.[16,28] Thus for fractional diffusion, based on the fractional derivative, it may be necessary to use the MLPPD rather than GPD to analyze the PFG fractional diffusion.

In this paper, from the MLPPD approximation, the general PFG signal attenuation expression is derived, which includes the finite gradient pulse width effect and works for all three types of fractional diffusions. Though this is a non-Gaussian phase approximation method, many results from the conventional GPD approximation method, such as Kärger et al.'s time-correlation function calculation results,[16] were employed. Therefore, the MLPPD method could be viewed as a modified GPD method. The results obtained agree with alternate derivations in the literature[23,28,31] and provide a set of convenient formalisms to interpret somewhat complicated PFG fractional diffusion behavior.

## II. THEORY

### A. Mainardi-Luchko-Pagnini phase distribution approximation

The fractional diffusion equation[14,15,30] based on fractional derivatives has been used to model anomalous diffusion in real space, which is

$$_t D_*^\alpha P(z,t) = D_f \,_z D_\theta^\beta P(z,t), \tag{1}$$

where $_\tau D_*^\alpha$ is the Caputo fractional derivative (see Appendix A), $_z D_\theta^{\beta'}$ is the Riesz-Feller space fractional derivative (see Appendix A), $z$ is the position, $D_f$ is the general fractional diffusion coefficient with units of m$^\beta$/s$^\alpha$ and $0 < \alpha, \beta \leq 2$.[14,15,30] By solving Eq. (1), the PDF[14,15,28] is given by

$$P(z,t) = \frac{1}{(D_f t^\alpha)^{1/\beta}} \kappa_{\beta',\alpha}^\theta \left( \frac{z}{(D_f t^\alpha)^{1/\beta}} \right). \tag{2}$$

The mean $\eta$-th power of the displacement can be calculated as[15,32]

$$\langle |z^\eta(t)| \rangle = M_{\beta,\alpha}(\eta)\left[Dt^\alpha\right]^{\frac{\eta}{\beta}}, \tag{3}$$

with

$$M_{\beta,\alpha}(\eta) = 2\int_0^\infty \varepsilon^\eta \kappa_{\beta,\alpha}^\theta(\varepsilon)d\varepsilon, \tag{4}$$

where $\varepsilon = z/(D_f t^\alpha)^{1/\beta}$.

In PFG experiments, the accumulated phase shift (the net non-refocused phase), $\phi(t)$, of the spin moments resulting from gradient pulses in a rotating frame can be described as[33,34,35]

$$\phi(t) = \int_0^t \gamma g(t')z(t')dt', \tag{5}$$

where $\gamma$ is the gyromagnetic ratio, $g(t')$ is the gradient strength at time $t'$, and position $z(t')$ is a time-dependent position related to the spin self-diffusion process. Typical PFG experiments consist of dephasing gradient pulses applied at the beginning of diffusion delay and rephasing gradient pulses applied after diffusion delay. Based on Eq. (5), if the spin carriers are immobile, $z(t')$ is a constant, and the accumulating phase will be refocused; while if the spin carriers are diffusing, the phase will not be refocused which leads to the signal attenuation. In the effective phase shift diffusion equation method, the APS can be described by a random walk model,[27,36] $\phi = \sum_{\Delta t}\Delta\phi = -\sum_{\Delta t}K(t)\Delta z(t)$, where $\Delta\phi$ is the phase jump length in the virtual phase space at time $t$, $K(t)$ is the wavenumber defined by $K(t) = \int_0^t \gamma g(t')dt'$ and $\Delta z(t)$ is the jump in real space at time $t$. Such a random jump process in a virtual phase space is similar to the diffusion process $\sum_{\Delta t}\Delta z(t)$. The difference between the two diffusion processes is the scale factor $K(t)$. Therefore, both diffusions belong to the same type of diffusion and they can be described by the same type of diffusion equation. The diffusion coefficient for the virtual phase diffusion process $-\sum_{\Delta t}K(t)\Delta z(t)$ can be expressed as[28]

$$D_{\phi eff}(t) = K^\beta(t)D_f, \tag{6}$$

where $D_{\phi eff}(t)$ is the effective phase shift diffusion coefficient with units of rad$^\beta$/s$^\alpha$. Based on eq. (1), by replacing the diffusion coefficient $D_f$ with $D_{\phi eff}(t)$, and replacing $z$ with $\phi$, the phase shift diffusion process can be described by the following equation

$$_tD_*^\alpha P(\phi,t) = D_{\phi eff}(t)_\phi D_\theta^\beta P(\phi,t),  \tag{7}$$

Using SGP approximation, the obtained phase shift is[28]

$$P(\phi,\Delta) = \frac{1}{\left[\langle|\phi^\beta(\Delta)|\rangle/M_{\beta,\alpha}(\beta)\right]^{1/\beta}} \kappa_{\beta',\alpha}^\theta\left(\frac{\phi}{\left[\langle|\phi^\beta(\Delta)|\rangle/M_{\beta,\alpha}(\beta)\right]^{1/\beta}}\right), \tag{8}$$

where $\Delta$ is the diffusion delay, and $\langle|\phi^\beta(\Delta)|\rangle$ is defined by [13,29]

$$\langle|\phi^\eta(\Delta)|\rangle = M_{\beta,\alpha}(\eta)\left[D_{\phi eff}(\delta)\Delta^\alpha\right]^{\frac{\eta}{\beta}}, \tag{9}$$

where $\delta$ is the gradient pulse length and $M_{\beta,\alpha}(\eta)$ is defined by Eq. (4). From Eqs. (2) and (8), it is obvious that both the PDFs of a particle in real space and in virtual phase space are Mainardi function distributions. The APS distribution including the finite gradient pulse width effect may be approximately treated using the MLPPD, which is

$$P(\phi,t) = \frac{1}{\left[\langle|\phi^\beta(t)|\rangle/M_{\beta,\alpha}(\beta)\right]^{1/\beta}} \kappa_{\beta',\alpha}^\theta\left(\frac{\phi}{\left[\langle|\phi^\beta(t)|\rangle/M_{\beta,\alpha}(\beta)\right]^{1/\beta}}\right). \tag{10}$$

If the value of $\langle|\phi^\beta(t)|\rangle/M_{\beta,\alpha}(\beta)$ is known, the PFG signal attenuation which resulted from non-refocused phase shift due to diffusion can be calculated by spatially averaging over all possible accumulating phase distribution as[28]

$$\begin{aligned}A(t) &= \frac{S(t)}{S(0)} \\ &= \int_{-\infty}^{+\infty} P(\phi,t)\cos(\phi)d\phi \\ &= E_\alpha\left[-\langle\phi^\beta(t)\rangle/M_{\beta,\alpha}(\beta)\right]\end{aligned}, \tag{11}$$

where $s(0)$ and $s(t)$ are signal intensities at the beginning and time t, respectively. The direct calculation of $\langle\phi^\beta(t)\rangle/M_{\beta,\alpha}(\beta)$ is difficult, however, based on Eq. (9), it is easy to obtain

$$\frac{\langle|\phi^\beta(t)|\rangle}{M_{\beta,\alpha}(\beta)} = \left[\frac{\langle\phi^2(t)\rangle}{M_{\beta,\alpha}(2)}\right]^{\frac{\beta}{2}}. \tag{12}$$

Using Eq. (5), $\langle \phi^2(t) \rangle / M_{\beta,\alpha}(2)$ can be defined as

$$\frac{\langle \phi^2(t) \rangle}{M_{\beta,\alpha}(2)} = \frac{1}{M_{\beta,\alpha}(2)} \left[ \int_0^t \gamma g(t') z(t') dt' \right]^2. \tag{13}$$

For the pulsed gradient spin echo (PGSE) and the pulsed gradient stimulated-echo (PGSTE) experiments, as shown in Figure 1, with constant gradient $g$, Eq. (13) can be calculated as[16,34]

$$\frac{\langle \phi^2(t) \rangle}{M_{\beta,\alpha}(2)} = \frac{\gamma^2 g^2}{M_{\beta,\alpha}(2)} \left[ \int_0^\delta \int_0^\delta z(t')z(t'')dt'dt'' + \int_\Delta^{\Delta+\delta} \int_\Delta^{\Delta+\delta} z(t')z(t'')dt'dt'' - 2\int_0^\delta \int_\Delta^{\Delta+\delta} z(t')z(t'')dt'dt'' \right]. \tag{14}$$

From Eq. (3), $\langle z^2(t) \rangle$ can be derived as

$$\langle z^2(t) \rangle = M_{\beta,\alpha}(2)\left[ D_f t^\alpha \right]^{\frac{2}{\beta}} = M_{\beta,\alpha}(2) D_f^{\frac{2}{\beta}} t^{\frac{2\alpha}{\beta}}. \tag{15}$$

Eq. (15) can be further rewritten as

$$\langle z^2(t) \rangle = M_{\beta,\alpha}(2)\left[ D_f t^\alpha \right]^{\frac{2}{\beta}} = M_{\beta,\alpha}(2) D_f^{\frac{2}{\beta}} t^\nu, \tag{16}$$

where $\nu = 2\alpha/\beta$. The similar calculation of Eq. (14) including the time correlation function $z(t')z(t'')$ has been derived by Kärger et al. in reference.[16] Based on Kärger et al.'s result, $\langle \phi^2(t) \rangle / M_{\beta,\alpha}(2)$ can be written as

$$\frac{\langle \phi^2(t) \rangle}{M_{\beta,\alpha}(2)} = \frac{2\gamma^2 g^2 D_f^{\frac{2}{\beta}}}{(\nu+1)(\nu+2)} \left[ \frac{1}{2}(\Delta+\delta)^{\nu+2} + \frac{1}{2}(\Delta-\delta)^{\nu+2} - \Delta^{\nu+2} - \delta^{\nu+2} \right]. \tag{17}$$

By substituting Eq. (17) into Eq. (12), we have

$$\frac{\langle |\phi^\beta(t)| \rangle}{M_{\beta,\alpha}(\beta)} = \gamma^\beta g^\beta D_f \left\{ \frac{2}{(\frac{2\alpha}{\beta}+1)(\frac{2\alpha}{\beta}+2)} \left[ \frac{1}{2}(\Delta+\delta)^{\frac{2\alpha}{\beta}+2} + \frac{1}{2}(\Delta-\delta)^{\frac{2\alpha}{\beta}+2} - \Delta^{\frac{2\alpha}{\beta}+2} - \delta^{\frac{2\alpha}{\beta}+2} \right] \right\}^{\frac{\beta}{2}}. \tag{18}$$

By substituting Eq. (18) into Eq. (11), we obtain

$$A(t) = E_\alpha \left[ -D_f b^*_{\alpha,\beta} \right]. \tag{19}$$

where $E_\alpha$ is the Mittag-Leffler function (see Appendix B) and

$$b^*_{\alpha,\beta} = \gamma^\beta g^\beta \left[ \frac{2}{(\frac{2\alpha}{\beta}+1)(\frac{2\alpha}{\beta}+2)} \left[ \frac{1}{2}(\Delta+\delta)^{\frac{2\alpha}{\beta}+2} + \frac{1}{2}(\Delta-\delta)^{\frac{2\alpha}{\beta}+2} - \Delta^{\frac{2\alpha}{\beta}+2} - \delta^{\frac{2\alpha}{\beta}+2} \right] \right]^{\frac{\beta}{2}}$$

or

$$b^*_{\alpha,\beta} = \begin{cases} \gamma^\beta g^\beta \delta^{\alpha+\beta} \left[ \frac{4(2^{\frac{2\alpha}{\beta}}-1)}{(\frac{2\alpha}{\beta}+1)(\frac{2\alpha}{\beta}+2)} \right]^{\frac{\beta}{2}} , \Delta = \delta \\ \gamma^\beta g^\beta \delta^\beta \Delta^\alpha , \delta << \Delta \\ \frac{2\gamma^2 g^2}{(\alpha+1)(\alpha+2)} \left[ \frac{1}{2}(\Delta+\delta)^{\alpha+2} + \frac{1}{2}(\Delta-\delta)^{\alpha+2} - \Delta^{\alpha+2} - \delta^{\alpha+2} \right] , \beta = 2 \\ \gamma^2 g^2 \delta^2 \left[ \Delta - \frac{1}{3}\delta \right] , \alpha = 1, \beta = 2 \end{cases} \qquad (20)$$

From Eqs. (19) and (20), the results agree with effective phase shift diffusion equation method at SGP approximation.[28] When $\alpha = 1, \beta = 2$, the result reduces to a normal diffusion result.[33,34] At small attenuation,

$$A(t) = E_\alpha(-D_f b^*_{\alpha,\beta}) \approx \exp\left[-\frac{D_f b^*_{\alpha,\beta}}{\Gamma(1+\alpha)}\right]. \qquad (21)$$

When $\beta = 2$, Eq. (21) reduces to

$$A(t) = \exp\left\{-\frac{2D_f \gamma^2 g^2}{(\alpha+1)(\alpha+2)\Gamma(1+\alpha)} \left[ \frac{1}{2}(\Delta+\delta)^{\alpha+2} + \frac{1}{2}(\Delta-\delta)^{\alpha+2} - \Delta^{\alpha+2} - \delta^{\alpha+2} \right]\right\}, \qquad (22)$$

which agrees with Kärger et al.'s results.[16]

### B. How to determine $\alpha$ and $\beta$?

It is important to determine the derivative parameters $\alpha$ and $\beta$ as they have been used as potential biomarkers for MRI.[37] From Eq. (21), at small attenuation, when the gradient pulse length $\delta$ and the diffusion delay $\Delta$ are fixed, we have

$$\ln[\ln A(0) - \ln A(t)] = \ln[\ln S(0) - \ln S(t)] = \ln\left[\frac{\gamma^\beta D_f}{\Gamma(1+\alpha)} C_{\alpha,\beta}(\delta,\Delta)\right] + \beta \ln(g), \qquad (23)$$

where $A(0)$ equals to 1 because there is no signal attenuation at the beginning of the first gradient pulse, $S(0)$ and $S(t)$

are the signal intensities at time 0 and time t, respectively, and $C_{\alpha,\beta}(\delta,\Delta)$ is

$$C_{\alpha,\beta}(\delta,\Delta) = \left[\frac{2}{(\frac{2\alpha}{\beta}+1)(\frac{2\alpha}{\beta}+2)}\left[\frac{1}{2}(\Delta+\delta)^{\frac{2\alpha}{\beta}+2} + \frac{1}{2}(\Delta-\delta)^{\frac{2\alpha}{\beta}+2} - \Delta^{\frac{2\alpha}{\beta}+2} - \delta^{\frac{2\alpha}{\beta}+2}\right]\right]^{\frac{\beta}{2}}. \quad (24)$$

From Eq. (21), at small attenuation, we can also have

$$\ln[\ln A(0) - \ln A(t)] = \ln[\ln S(0) - \ln S(t)] = \begin{cases} \ln\left[\frac{\gamma^\beta g^\beta D_f C_2}{\Gamma(1+\alpha)}\right] + (\alpha+\beta)\ln(\delta), & \text{when } \Delta = \delta \\ \ln\left[\frac{\gamma^\beta g^\beta D_f}{\Gamma(1+\alpha)}\right] + \alpha\ln(\Delta), & \text{when } \delta << \Delta \end{cases}, \quad (25)$$

where $C_2$ is

$$C_2 = \left[\frac{4\delta^{\frac{2\alpha}{\beta}+2}}{(\frac{2\alpha}{\beta}+1)(\frac{2\alpha}{\beta}+2)}\left[2^{\frac{2\alpha}{\beta}+1} - 1\right]\right]^{\frac{\beta}{2}}. \quad (26)$$

Similar equations as Eqs. (23) and (25) have been obtained by instantaneous signal attenuation (ISA) method.[31] Eqs. (23) and (25) may be used to determine the fractional time and space derivative parameters $\alpha$ and $\beta$ in PFG fractional diffusion experiments.

### III. DISCUSSION

The general accumulating phase shift distribution and signal attenuation expressions for fractional diffusion were obtained based on the MLPPD approximation. The obtained signal attenuation expression $E_\alpha[-D_f b^*_{\alpha,\beta}]$ is consistent with other theoretical results in the literature.[23,28,31] For space-fractional diffusion, $\{\alpha=1, 0<\beta\leq 2\}$, as $E_\alpha(-x) = \exp(-x)$ when $\alpha=1$, the signal attenuation expression Eq. (19) can be rewritten as

$$A(t) = \exp\left[-\gamma^\beta g^\beta D_f C_{1,\beta}(\delta,\Delta)\right], \quad (27)$$

where $C_{1,\beta}(\delta,\Delta)$ is

$$C_{1,\beta}(\delta,\Delta) = \left[\frac{2}{(\frac{2}{\beta}+1)(\frac{2}{\beta}+2)}\left[\frac{1}{2}(\Delta+\delta)^{\frac{2}{\beta}+2} + \frac{1}{2}(\Delta-\delta)^{\frac{2}{\beta}+2} - \Delta^{\frac{2}{\beta}+2} - \delta^{\frac{2}{\beta}+2}\right]\right]^{\frac{\beta}{2}}. \quad (28)$$

Eq. (27) is close to the theoretical result

$$A(t) = \exp\left[-\gamma^\beta g^\beta \delta^\beta (\Delta - \frac{\beta-1}{\beta+1}\delta)\right], \tag{29}$$

obtained by the modified Bloch equation,[23] the effective phase shift diffusion equation method,[28] and the instantaneous signal attenuation method.[31] Figure 2 shows the comparison of $C_{1,\beta}(\delta,\Delta)$ in Eqs. (27) and $\Delta - \frac{\beta-1}{\beta+1}\delta$ in Eq. (29). The differences between the MLPPD approximation and other methods are small. The greater the value of $\beta$ is, the smaller the difference is, since at $\beta = 2$, both Eqs. (27) and (29) reduce to normal diffusion based attenuation. Additionally, the greater the ratio $\delta/\Delta$ is, the smaller the difference between methods is, since under SGP approximation, both Eqs. (27) and (29) are reduced to $\exp(-\gamma^\beta g^\beta \delta^\beta \Delta)$.

Furthermore, the signal attenuation from MLPPD approximation for time-fractional diffusion agrees with that obtained by the instantaneous signal attenuation method.[31] When $0 < \alpha \leq 2, \beta = 2$, the signal attenuation expression Eq. (19) can be written as

$$A(t) = E_\alpha\left\{-D_f \gamma^2 g^2 C_{\alpha,2}(\delta,\Delta)\right\}, \tag{30}$$

where $C_{\alpha,2}(\delta,\Delta)$ is

$$C_{\alpha,2}(\delta,\Delta) = \frac{2}{(\alpha+1)(\alpha+2)}\left[\frac{1}{2}(\Delta+\delta)^{\alpha+2} + \frac{1}{2}(\Delta-\delta)^{\alpha+2} - \Delta^{\alpha+2} - \delta^{\alpha+2}\right]. \tag{31}$$

Eq. (30) is close to the signal attenuation obtained by the instantaneous signal attenuation method,[27] which is

$$A(t) = E_\alpha\left\{-D_f \gamma^2 g^2 C'_{\alpha,2}(\delta,\Delta)\right\}, \tag{32}$$

where $C'_{\alpha,2}(\delta,\Delta)$ is[26, 28]

$$C'_{\alpha,2}(\delta,\Delta) = \gamma^2 g^2 \left\{\frac{\alpha \delta^{\alpha+2}}{\alpha+2} + \delta^2(\Delta^\alpha - \delta^\alpha) + \frac{2}{(\alpha+1)(\alpha+2)}\left[(\Delta+\delta)^{2+\alpha} - \Delta^{2+\alpha}\right] - \frac{2}{(\alpha+1)}\Delta^{1+\alpha}\delta - \Delta^\alpha \delta^2\right\}. \tag{33}$$

The comparison of $C_{\alpha,2}(\delta,\Delta)$ and $C'_{\alpha,2}(\delta,\Delta)$ is shown in Figure 3. The difference between the MLPPD approximation and the instantaneous signal attenuation method is small for $0.5 < \alpha < 1.5$. At $\alpha = 1$, there is no difference because both Eqs. (30) and (32) reduce to normal diffusion. Similarly, the smaller the ratio $\delta/\Delta$ is, the smaller the difference is, since again under SGP approximation, both Eqs. (30) and (32) can reduce to $E_\alpha(-\gamma^2 g^2 \delta^2 \Delta^\alpha)$.

Contrasting the time-fractional diffusion result with the space-fractional diffusion result shows that the deviations from other methods were greater with variation in the time derivative parameter $\alpha$ than with the variation in the space derivative parameter $\beta$. For the general attenuation, the instantaneous signal attenuation method gives $E_\alpha\left[-D_f b'^*{}_{\alpha,\beta}\right]$, where $b'^*{}_{\alpha,\beta} = \int_0^t K^\beta(t) dt^\alpha$ ($K(t)$ is the wavenumber). Similarly, it should be expected that the MLPPD result is close to the instantaneous signal attenuation result at smaller $\delta/\Delta$ ratios, because, under the SGP approximation, both methods lead to the same attenuation expression $E_\alpha\left(-\gamma^\beta g^\beta \delta^\beta \Delta^\alpha\right)$. Additionally, at $\alpha=1, \beta=2$, both signal attenuation expressions reduce to the signal attenuation result of restricted normal diffusion.

The derivative parameters $\alpha$ and $\beta$ can be determined by Eqs. (23) and (25). Particularly, in PFG fractional diffusion experiments, the $\beta$ may be determined at first by Eq. (23), with the parameters $\alpha$ and $D_f$ being the two only unknown parameters in the experiments, which leads to a less arbitrary interpretation of experimental data with fewer floating parameters.

The MLPPD approximation provides a set of convenient formalisms for PFG fractional diffusion. The MLPPD method used is a modified GPD method. Considering the broad application of the GPD method in normal diffusion, the MLPPD method may also have a broad application for various PFG fractional diffusion problems, such as restricted fractional diffusion [38] including the finite gradient pulse width effect.

**APPENDIX A. DEFINITION OF FRACTIONAL DERIVATIVE**[14]

Caputo fractional derivative in time[14]

$$_tD_*^\alpha f(t) := \begin{cases} \dfrac{1}{\Gamma(m-\alpha)} \int_0^t \dfrac{f^{(m)}(\tau)d\tau}{(t-\tau)^{\alpha+1-m}}, & m-1 < \alpha < m, \\ \dfrac{d^m}{dt^m} f(t), & \alpha = m. \end{cases} \quad (A1)$$

Riesz-Feller derivative in space for $0 < \beta < 2$ and $|\theta| \leq \min\{\beta, 2-\beta\}$[12]

$$_xD_\theta^\beta f(x) = \frac{\Gamma(1+\beta)}{\pi}\left\{\sin[(\beta+\theta)\pi/2]\int_0^\infty \frac{f(x+\xi)-f(x)}{\xi^{1+\beta}} d\xi \right. \\ \left. +\sin[(\beta-\theta)\pi/2]\int_0^\infty \frac{f(x-\xi)-f(x)}{\xi^{1+\beta}} d\xi\right\}. \quad (A2)$$

# APPENDIX B. DEFINITION OF $\kappa_{\beta',\alpha}^{\theta}(x)$ FUNCTION,[14] MITTAG-LEFFLER FUNCTION $E_{\alpha,1}(x)$ [14]

$$\kappa_{\beta',\alpha}^{\theta}(x) = \frac{1}{\beta'x} \frac{1}{2\pi i} \int_{\gamma-i\infty}^{\gamma+i\infty} \frac{\Gamma(\frac{s}{\beta'})\Gamma(1-\frac{s}{\beta'})\Gamma(1-s)}{\Gamma(1-\frac{\alpha}{\beta'}s)\Gamma(\rho s)\Gamma(1-\rho s)} x^s ds, \text{with} \quad \rho = \frac{\beta'-\theta}{2\beta'}. \tag{B1}$$

$$E_{\alpha,1}(-x) = \sum_{n=0}^{\infty} \frac{(-x)^n}{\Gamma(n\alpha+1)} = \begin{cases} \exp(-x), \alpha = 1 \\ \approx \exp\left(-\frac{x}{\Gamma(1+\alpha)}\right), |x| \ll 1 \end{cases}. \tag{B2}$$

**Figure Legends**

**FIG. 1** (a) PGSE pulse sequence with gradient pulses of finite length $\delta$ and diffusion delay $\Delta$, (b) PGSTE pulse sequence with gradient pulses of finite length $\delta$ and diffusion delay $\Delta$.

**FIG. 2** Comparison $C_{1,\beta}(\delta,\Delta)$ in the space-fractional diffusion signal attenuation expression of Eqs. (27) from MLPPD approximation with $\Delta - \frac{\beta-1}{\beta+1}\delta$ in Eq. (29) from modified Bloch equation,[23] effective phase shift diffusion equation[26] and instantaneous signal attenuation methods.[31]

**FIG. 3** Comparison of $C_{\alpha,2}(\delta,\Delta)$ in time-fractional diffusion signal attenuation expression Eq. (30) from MLPPD approximation with $C'_{\alpha,2}(\delta,\Delta)$ in Eq. (32) from instantaneous signal attenuation methods.[31]

**FIG. 1**

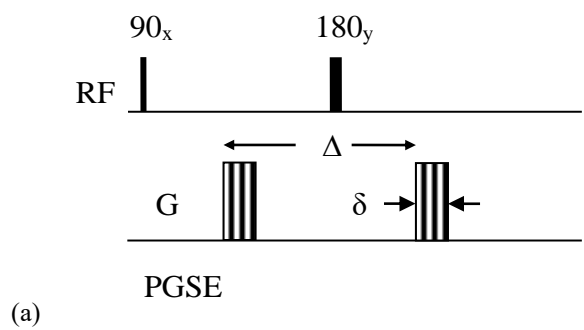

(a) PGSE

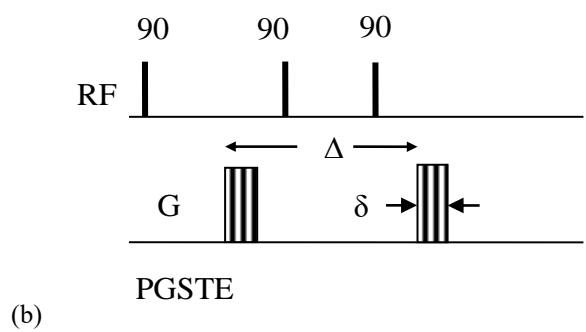

(b) PGSTE

**FIG. 2**

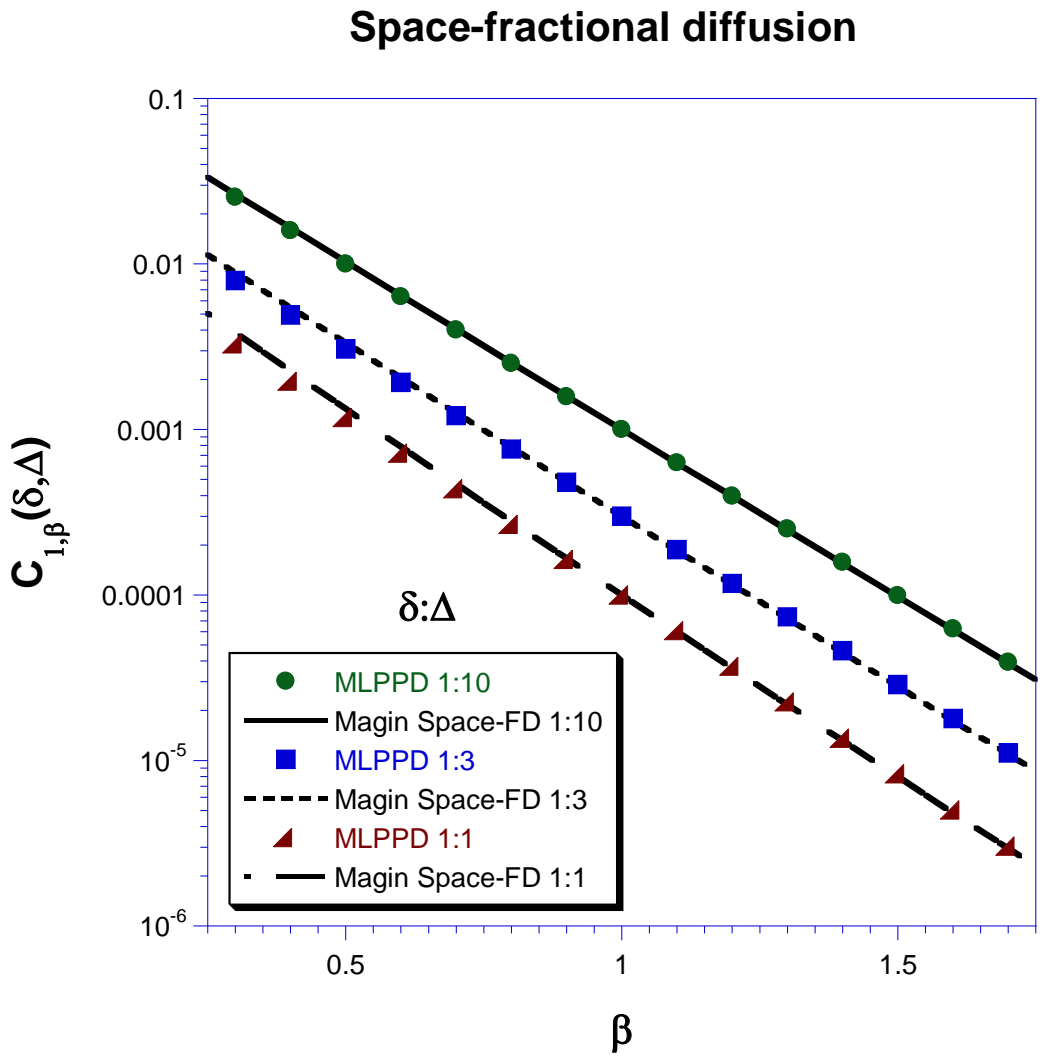

**FIG. 3**

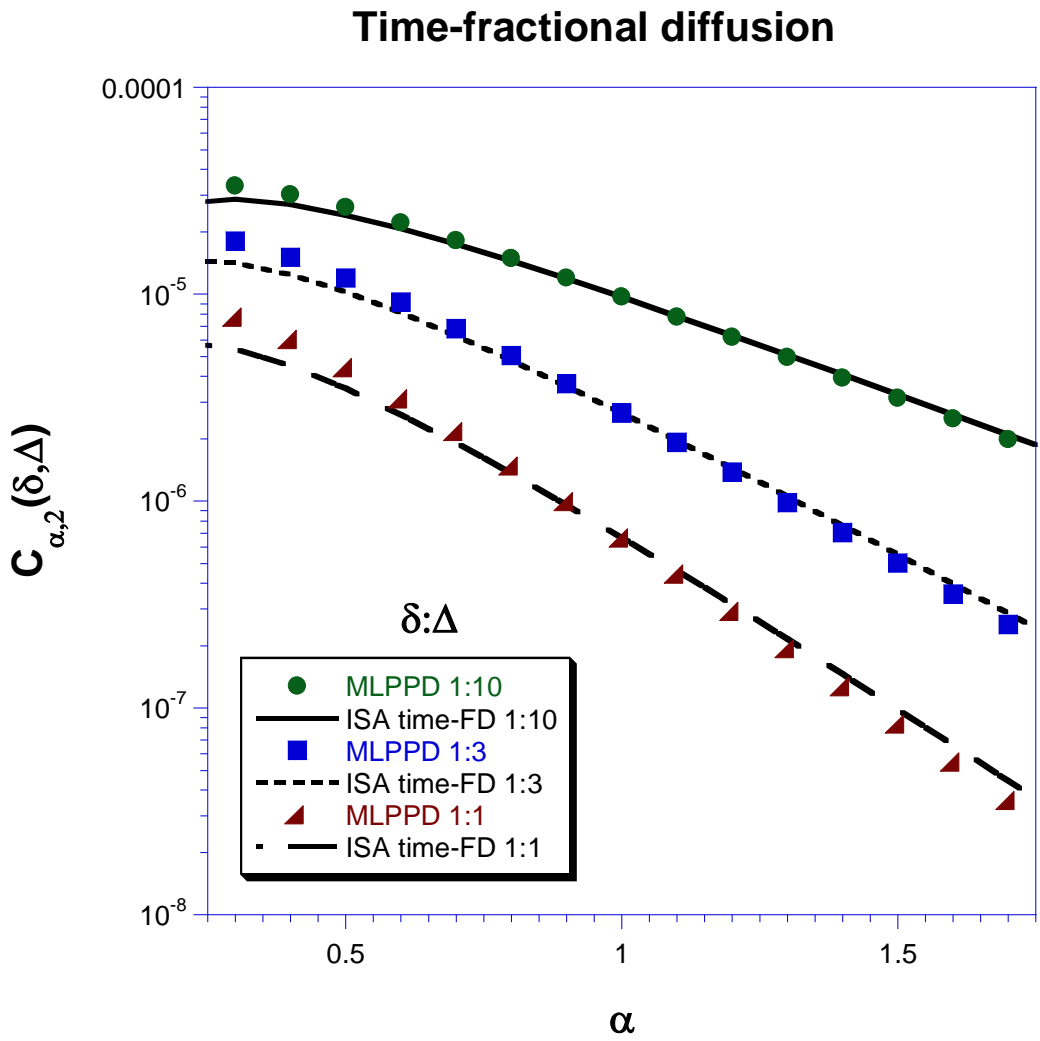